\documentclass[global,final]{svjour}
\usepackage{graphics}
\usepackage{fixltx2e}
\graphicspath{{images/},{figures/}}
\journalname{Appl. Phys. B}
\begin{document}
\title{Design of the $10\,\rm{m}$ AEI prototype facility\\ for interferometry studies}
\subtitle{A brief overview}
\author{T Westphal\inst{1} \and G Bergmann\inst{1} \and A Bertolini\inst{1} \and M Born\inst{1} \and Y Chen\inst{2} \and A V Cumming\inst{3} \and L Cunningham\inst{3}  \and K Dahl\inst{1} \and
C Gr\"af\inst{1} \and G Hammond\inst{3} \and G Heinzel\inst{1} \and S Hild\inst{3} \and S Huttner\inst{3} \and R Jones\inst{3} \and F Kawazoe\inst{1} \and S K\"ohlenbeck\inst{1} \and G K\"uhn\inst{1} \and H L\"uck\inst{1} \and K Mossavi\inst{1} \and J H P\"old\inst{1} \and K Somiya\inst{4} \and A M van Veggel\inst{3} \and A Wanner\inst{1} \and B Willke\inst{1} \and K A Strain\inst{3} \and S Go\ss ler\inst{1} \and K Danzmann\inst{1}
\thanks{\emph{Present address:} Tobias.Westphal@aei.mpg.de}%
}                     
\offprints{Tobias Westphal}          
\institute{Max-Planck-Institut f\"ur Gravitationsphysik (AEI), Leibniz Universit\"at Hannover, D-30167 Hannover, Germany \and California Institute of Technology, Theoretical Astrophysics 130-33, Pasadena, CA 91125, USA \and SUPA, University of Glasgow, Glasgow, G12 8QQ, UK \and Tokyo Institute of Technology
2-12-1 Ookayama, Meguro-ku, Tokyo 152-8550, Japan}
\date{ }
%
\maketitle
\begin{abstract}
The AEI 10\,m prototype interferometer facility is currently being constructed at the Albert Einstein Institute in Hannover, Germany. It aims to perform experiments for future gravitational wave detectors using advanced techniques. Seismically isolated benches are planned to be interferometrically interconnected and stabilized, forming a low-noise testbed inside a $100\,\rm{m}^3$ ultra-high vacuum system. A well-stabilized high power laser will perform differential position readout of $100\,\rm{g}$ test masses in a $10\,\rm{m}$ suspended arm-cavity enhanced Michelson interferometer at the crossover of measurement (shot) noise and backaction (quantum radiation pressure) noise, the so-called Standard Quantum Limit (SQL). Such a sensitivity enables experiments in the highly topical field of macroscopic quantum mechanics. In this article we introduce the experimental facility and describe the methods employed, technical details of subsystems will be covered in future papers.
\end{abstract}
\section{Introduction}
Gravitational wave detectors are the most sensitive differential length change detectors in the world~\cite{LIGO,Virgo,GEO600}. Yet their incredible sensitivity is not available for quantum mechanical experiments as they are built for their dedicated purpose of listening to the universe. There are several interferometer prototype facilities around the world \cite{glasgow10m,caltech40m,lasti10m,gingin80m} with the main purpose of further increasing these detectors' sensitivities. The defining feature of the system described in this article is that it is focused on both fundamental questions in quantum mechanics and technical performance enhancements. To meet this challenge a $10\,\rm{m}$ prototype interferometer facility is under construction in Hannover, Germany. A $100\,\rm{m}^3$ ultra-high vacuum (UHV) system provides space for several simultaneous experiments (sec. \ref{sec:vac}). It has been designed for easy access and rapid pump down. 

A virtual pre-isolated platform is formed from three separate optical tables (sec. \ref{sec:tables}). The distances between these, and their relative angles are sensed interferometrically, and two of them are forced to follow the third to a precision far better than the thermal expansion of a single large table in a conventional temperature stabilized lab (sec. \ref{sec:SPI}). This removes the dominant disturbance of interferometric experiments to be performed on these tables on time-scales of hours or longer.  
At higher frequencies, around the microseismic peak (at about $0.15\,\rm{Hz}$), the relative motion between the tables is suppressed by three to four orders of magnitude. This virtual platform is probably the quietest place on Earth in terms of local relative motion. For this reason it is ideally suited for testing subsystems of space missions relying on free falling test masses such as Grace follow-on \cite{Gracefollowon} and LISA type missions \cite{LISA,lisa}. Furthermore it allows to realize a highly stable frequency reference for the powerful ($38\,\rm{W}$), highly amplitude stabilized Nd:YAG laser (sec. \ref{sec:laser}).

The first major goal is to design and build an apparatus able to reach the standard quantum limit (SQL) for a system with macroscopic ($100\,\rm{g}$) mirrors (sec. \ref{sec:SQLifo}). Such a system is limited in sensitivity by quantum noise in a wide band around the frequency at which shot noise (the measurement noise) and radiation pressure noise (the back-action noise) are equal. The margin between the sum of the classical noise contributions and the SQL is expected to be at least a factor of 2.5.  This will enable research in quantum mechanics with macroscopic optical components and light fields. 

A range of experiments is made possible by a system where classical noise is insignificant.  These include the observation of ponderomotive squeezing~\cite{pondsqz} due to strong opto-mechanical coupling and the entanglement of macroscopic mirrors via light pressure~\cite{entanglement}. While many experiments are already employing optical cooling of oscillators to low occupation numbers in the hunt to reach the ground state~\cite{cooling,cooling2}, a macroscopic experiment can reveal underlying decoherence processes such as gravity self decoherence \cite{decoherence}. Furthermore, nonclassical interferometry, e.g. frequency-dependent squeezed light injection or back-action evasion readout, can be investigated~\cite{QND}.

\section{Vacuum system}
\label{sec:vac}
\begin{figure*}
\resizebox{0.75\textwidth}{!}{
  \includegraphics{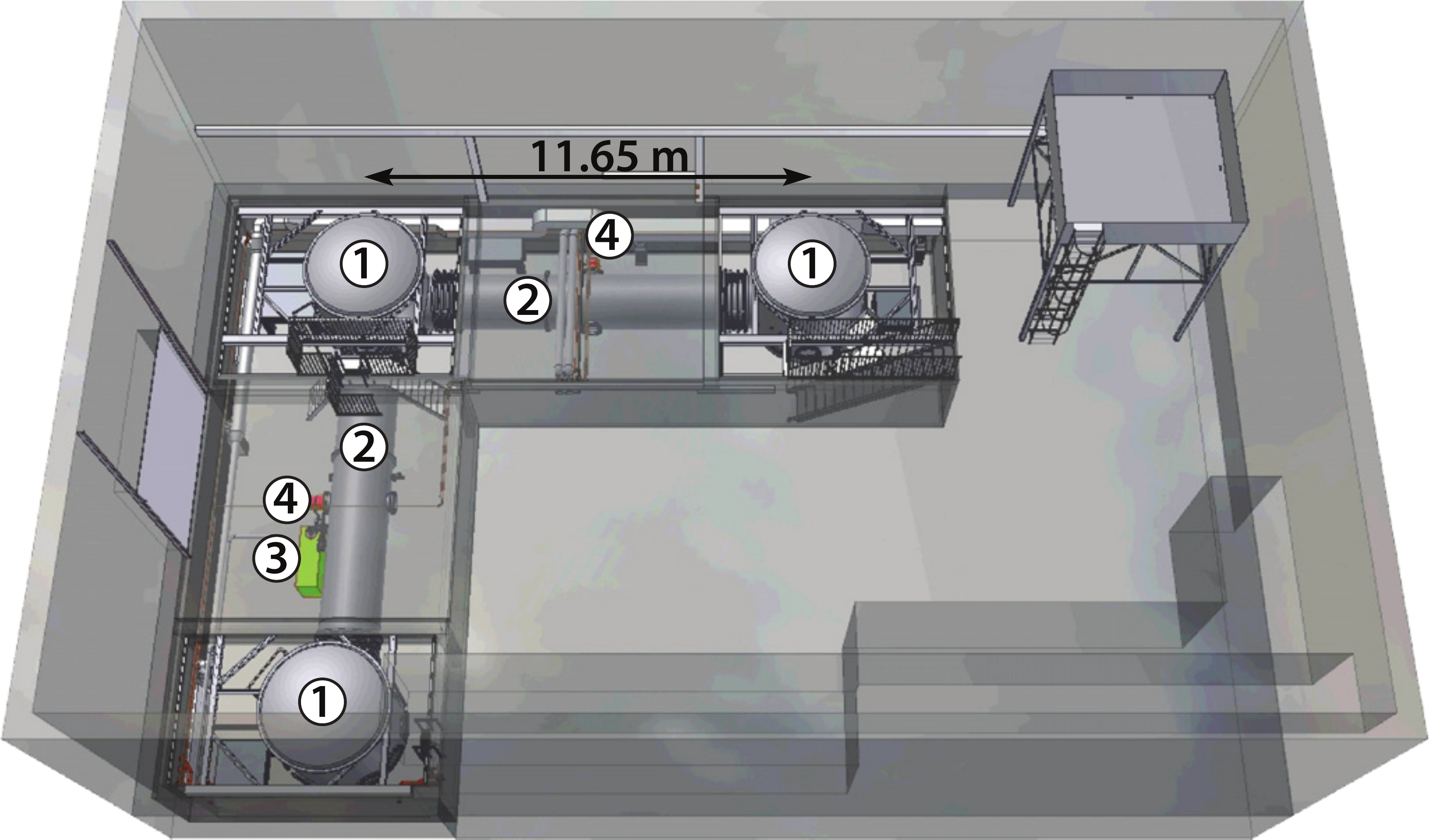}
}
\vspace*{0.5cm}    
\caption{The vacuum system was designed to fit into the basement of the AEI prototype hall. Three "walk-in" sized tanks (1) are interconnected by $1.5\,\rm{m}$ diameter tubes(2) in an "L-shaped" configuration of $11.65\,\rm{m}$ arm length. Fast pump down is provided by a screw pump (3). A pressure of $5\times 10^{-8}\,\rm{hPa}$ is reached with two turbo-molecular pumps (4) attached to the middle of the arm tubes.}
\label{fig:PThall} 
\end{figure*}

To reduce the influence of air damping, of refractive index fluctuations and of acoustic coupling, a large ultra-high vacuum system is used. Three tanks of $3.4\,\rm{m}$ height and $3\,\rm{m}$ diameter are interconnected by $1.5\,\rm{m}$ diameter tubes to form an "L-shape" (see fig.~\ref{fig:PThall}). The center-center distance of the tanks (the arm length) is $11.65\,\rm{m}$. To facilitate the hardware installation, the system can be entered through 1\,m doors. The large-aperture beam tubes allow light beams to be sent from any point on a table to any point on its neighbor.

A screw pump ($175\,\rm{l/s}$) pumps the $100\,\rm{m^3}$ system from atmospheric pressure down to $5\,\rm{Pa}$ within two hours. Then, two turbo-molecular pumps ($2400\,\rm{l/s}$ each) attached to the center of each arm are switched on, pumping to $10^{-6}\,\rm{hPa}$, sufficient for most experiments, within 10 hours. A pressure of $\le10^{-7}\,\rm{hPa}$ is reached within one week. The turbo pumps are backed by one scroll pump. While flanges up to $600\,\rm{mm}$ are sealed by copper gaskets, bigger ones are sealed by double Viton\textsuperscript{\textregistered} O-rings with the gap in between being separately evacuated by another shared scroll pump to reduce the leak rate by differential pumping. Since these pumps are running during interferometer operation, they are located in a separate room and are seismically decoupled by triple stacks of Sorbothane\textsuperscript{\textregistered} hemispheres and granite plates of several hundred kg. A pressure of $\approx5\times10^{-8}\,\rm{hPa}$ was reached, limited by partial pressure of water, since the system cannot be baked out at high temperatures.

\section{Seismically isolated and interlinked tables}
\label{sec:tables}
\begin{figure*}
\resizebox{0.75\textwidth}{!}{
  \includegraphics{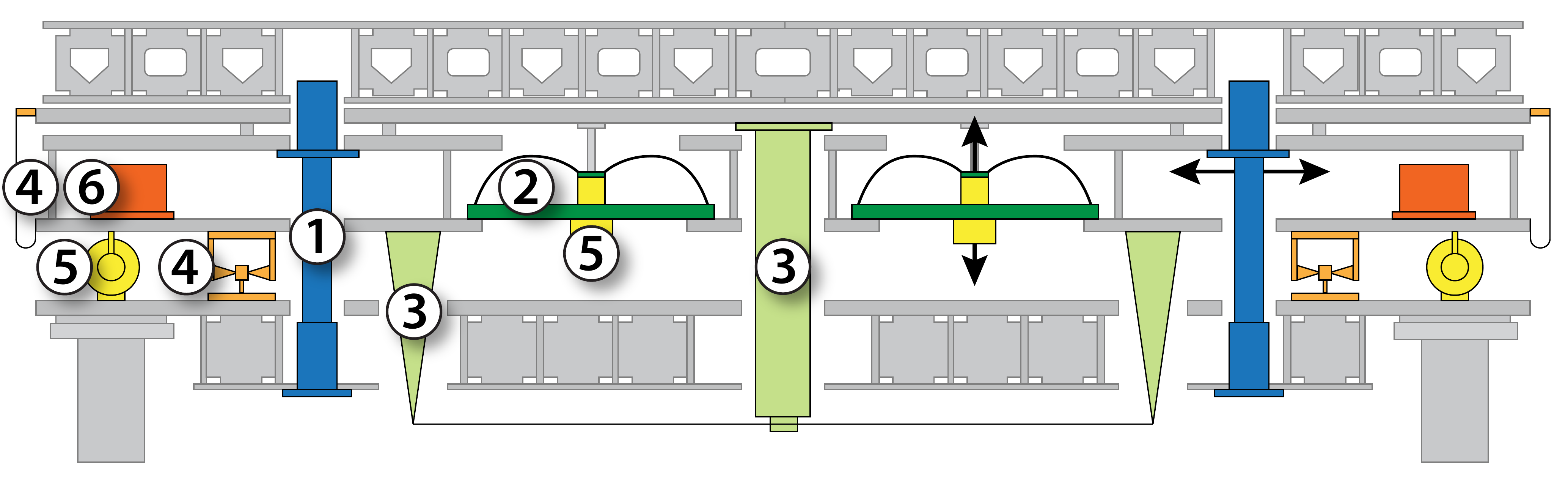}
}
\vspace*{0.5cm}    
\caption{Passively seismically isolated optical benches were developed on the basis of the HAM-SAS design~\cite{hamsas}. Inverted pendulum legs (1) provide horizontal isolation from seismic noise starting at very low frequencies ($0.05\,\rm{Hz}$). Vertical decoupling is provided by geometric anti-springs (2) electronically tuned to a resonance frequency of $0.1\,\rm{Hz}$ by means of positive feedback. The so-called tilt stabilization system (3) provides additional elastic restoring torque to the pitch and roll modes for the lowest frequency tunes of the vertical springs and/or highest positions of the optical table (including the payload) center of mass. Static position/attitude and thermal drifts are compensated by means of stepper motor driven springs (4). Linear variable differential transformers (5) and specifically designed accelerometers (6) sense the table's position and acceleration to produce feedback signals for active noise suppression and damping carried out by actuators co-located with the LVDTs.
}
\label{fig:SAS}
\end{figure*}

On a microscopic level the ground is continuously moving driven by natural and anthropogenic sources, so an isolated platform is needed for the experiments. A passively isolated optical bench ($1.75\,\rm{m}\times1.75\,\rm{m}$) are beeing installed in each of the three tanks (the central table is in place, the west table is close to installation). Local sensors as well as a Suspension Platform Interferometer are used for damping of the eigenmodes and providing additional active isolation at low frequencies. The mass of each optical table is $950\,\rm{kg}$, including $230\,\rm{kg}$ ballast, the most part of which can be converted into payload. Inside the vacuum tanks, each table is supported by a vibration isolator system (see fig.~\ref{fig:SAS}) providing passive seismic attenuation in all six degrees of freedom above the corresponding natural frequencies (all of them less than $200\,\rm{mHz}$).

The isolator system consists of a tilt rigid platform (spring box) mounted on three inverted pendulum (IP) legs providing seismic attenuation for horizontal translation and yaw. The anti-restoring force (exerted by gravity) is counteracted by flexures to give a stable design. The resonance frequency of the inverted pendulum is tuned down to $0.05\,\rm{Hz}$ by loading the spring box (adding ballast).

For the isolation along vertical, pitch and roll degrees of freedom, the spring box hosts three vertical spring systems (filters). To keep the tables compact, the main parts (IP and filters) are interleaved. Since vertical isolators cannot benefit from an anti-restoring gravitational potential, the geometric anti-spring (GAS) approach was chosen; cantilever springs are connected to a common center. There they are loaded strongly such that they bend. Compression of the blades towards this center creates a tunable anti-spring effect. A shallow potential can be obtained around the working point~\cite{GAS}. By adjusting the compression, the resonance frequency was tuned down to $0.25\,\rm{Hz}$. Further electronic tuning by means of positive feedback gave $0.1\,\rm{Hz}$. Pitch and roll isolation relies on the vertical and angular compliance of the GAS filters and on the position of the optical table's (including the payload) center of gravity (COG) located several centimeters above the GAS virtual pivot point. Additional elastic restoring torque allowing especially low frequency tunings of the GAS springs and/or highest COG positions is provided by a tilt stabilization device.
The non-zero moment of inertia of the IP legs and GAS blades would cause the transmissibility to saturate at $-60\,\rm{dB}$ in both cases ("center of percussion effect"). Suitable adjustable counterweight systems~\cite{IP},\cite{magicwand} have been implemented to achieve additional $20\,\rm{dB}$ improvement. 

Each isolator is instrumented with voice-coil linear actuators for the control of the optical table in all six degrees of freedom and with local sensors. The horizontal inertial motion of the spring box is measured by three custom designed UHV compatible accelerometers while three commercial L22 vertical geophones (Geospace Technologies are placed inside the optical tables (into suitable sealed vacuum cans) to monitor their pitch, roll and vertical movement. Above $1\,\rm{Hz}$ passive isolation provides subnanometer and subnanoradian residual root-mean-square motion, while at lower frequencies the system eigenmodes are actively controlled using local (linear variable differential transformers (LVTD) and accelerometers) and global (interferometric link) sensors. Below the microseismic peak, the residual differential motion between the tables would be too high for the planned experiments related to space missions such as LISA or Grace follow-on. Therefore, an additional readout system is being developed, as described in the next section.

\section{Suspension Platform Interferometer (SPI)}
\label{sec:SPI}
\begin{figure*}
\resizebox{0.75\textwidth}{!}{
  \includegraphics{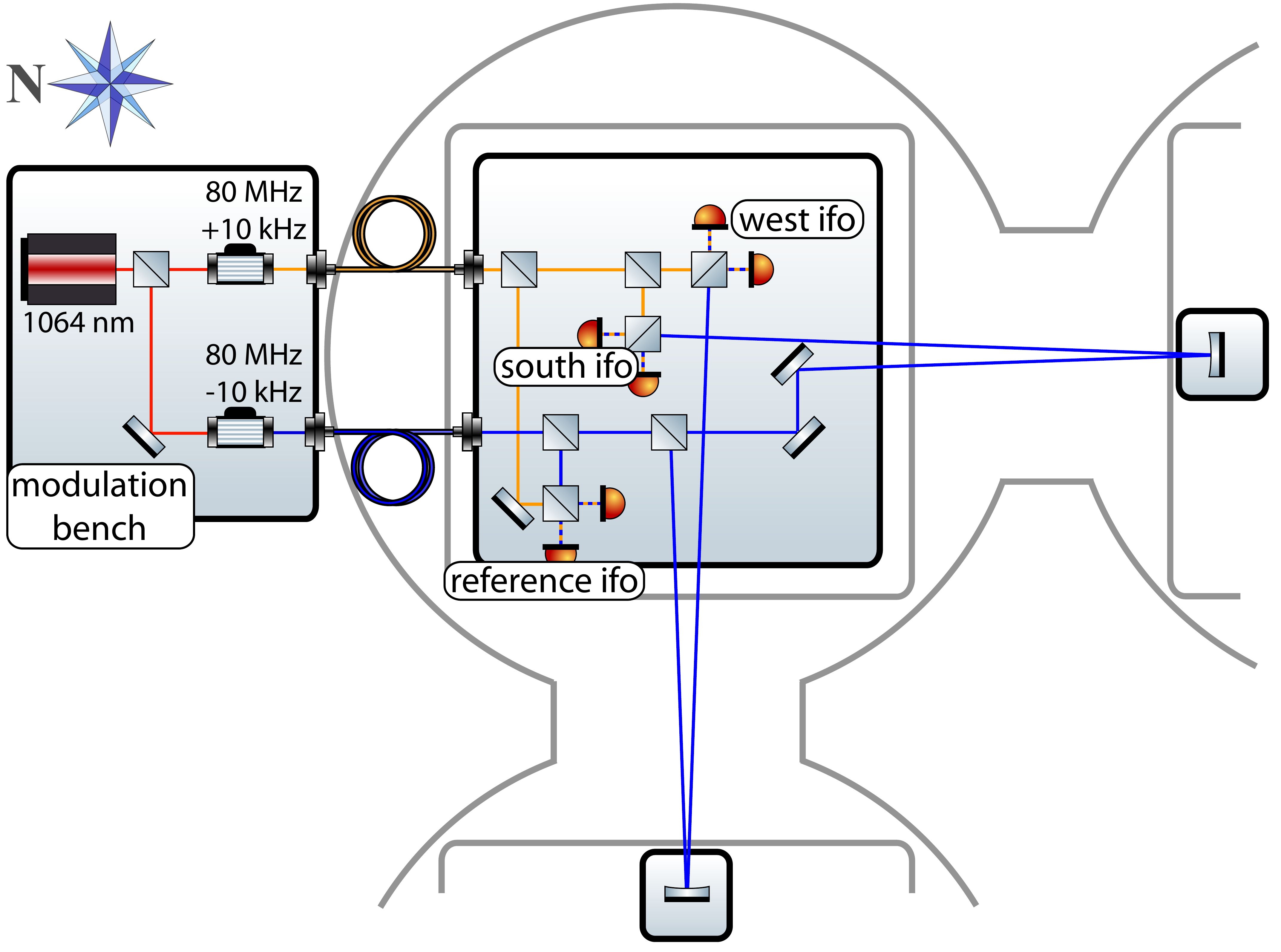}
}
\vspace*{0.5cm}      
\caption{The suspension platform interferometer (SPI) measures the relative position of the optical benches by means of heterodyne mach zehnder interferometers. While at high frequencies their passive isolation system delivers decoupling from ground motion, at low frequencies the SPI and local sensors signals are uses to actively stabilize the tables to give a single platform.}
\label{fig:SPI}    
\end{figure*}

The Suspension Platform Interferometer forms a virtual interferometric interconnection between the center table and each of the end tables (see fig.~\ref{fig:SPI}). The idea is to actively control the positions (and angles) of the tables in the low frequency range, where no passive isolation of the tables takes place. Thereby it is possible to maintain the center-center distance with a high degree of stability, thus creating a unique low-frequency displacement environment. The goal for the inter-table distance stability is $100\,\rm{pm/\sqrt{Hz}}$ between $10\,\rm{mHz}$ and $100\,\rm{Hz}$~\cite{SPI}. For pitch and yaw the goal is $10\,\rm{nrad/\sqrt{Hz}}$. Roll about the optical axis between two adjacent tables is not sensed. 

To carry out the SPI sensing, an Nd:YAG laser beam is split by a beamsplitter. The resulting beams get shifted by $80\,\rm{MHz}\pm 10\,\rm{kHz}$ by means of acousto optical modulators and are finally superimposed in a heterodyne Mach-Zehnder setup. One of them, the reference beam, is kept on a plate ($250\times250\times30\,\rm{mm^3}$) located in the center of the table in the central tank. To this plate, which is made of ultra low thermal expansion glass (Clearceram\textsuperscript{\textregistered}-Z\,HS), all optical components are attached by hydroxide catalysis bonding. The measurement beam, however, is split into four identical copies. Two of those are kept on the central plate. They are brought to interference with the reference beam in interferometers having identical path lengths. One of them, the reference interferometer, measures all differential phase delays introduced upstream, such as phase noise in the fibers feeding the laser beams into the vacuum system. The second, namely the diagnostic interferometer (not shown in fig.~\ref{fig:SPI} for simplicity), serves for debugging purposes and out-of-loop measurements. Another copy is sent to the west end table where it is reflected by a mirror (radius of curvature $-11.8\,\rm{m}$) placed in the middle of the table. The beam is reflected back to the central plate under a small angle. There it is interfered with the reference beam. An equivalent interferometer is formed by the fourth beam sensing the south table position. Such interferometers with $23\,\rm{m}$ arm length mismatch require the laser to be well frequency stabilized. For this reason the iodine reference option for the \textsl{Innolight Prometheus} laser was chosen \cite{iodine}. Quadrant photodiodes are used throughout the whole SPI. Thus, angular misalignments can be detected via differential wavefront sensing \cite{DWS}. An LTP style phase meter \cite{phasemeter} based on field programmable gate arrays (FPGAs) for fast data processing is employed for the photodiode readout and performs single bin discrete Fourier transforms on the signals. A micro controller processes the signals further, sending DC-power and heterodyne amplitude and phase data via Ethernet to the control and data system (see following section). There calibration and subtraction of the reference interferometer signal are carried out. Furthermore, channels are combined and digitally filtered to produce suitable actuation signals for the table position control.

A test setup has been built to verify the working principles. Currently the final SPI is waiting for the second table to be installed. Then the recombining beamsplitter of the first long arm (west interferometer) can be bonded in situ to finally demonstrate the full sensitivity and stabilize the corresponding relative table positions and angles.

\section{Control and data system (CDS)}
Digital control offers the flexibility that is needed for a prototype facility. 
To operate the $10\,\rm{m}$ prototype interferometer, many subsystems have to work together. This is orchestrated by a digital CDS that was developed at Caltech to operate the Advanced LIGO gravitational wave detectors [1]. The CDS front-end computers run more than hundred real-time control loops under a real-time enhanced Linux operating system to e.g. control mirror positions and laser beam parameters. The data of all involved signals can be stored to hard disks for later analysis of experimental results. Together with the data, a precise time stamp derived from a GPS synchronized clock is recorded.
The digital control loops are designed graphically from generic digital filters and Simulink\textsuperscript{\textregistered}-like blocks. A real-time code generator (RCG) compiles such a model into a Linux kernel module. The module is then assigned to a dedicated cpu core of a front-end computer to be able to run its control loops at up to $65\,\rm{kHz}$.
The digital filter coefficients, switches and parameters of the control loops can be changed online via a graphical user interface (MEDM screens) that sends commands over an EPICS~\cite{epics} channel access network protocol.
Each cpu core can run dozens of digital filters and the signals can be distributed between cpu cores and front-end computers.
The front-end computers are off-the-shelf Intel XEON and AMD Opteron servers. To digitize analog signals from the sensors, 16 bit ADC PCI-X cards\footnote{General Standards Corporation 16AI64SSC} with 32 differential channels each are housed in separate I/O-chassis. $16\,\rm{bit}$ DAC PCI-X cards\footnote{General Standards Corporation 16AO16} are located in the same chassis to drive the actuators, each with 16 differential output channels. The ADC and DAC cards are clocked externally with a GPS-locked $2^{16}\,\rm{Hz}$ ($65536\,\rm{Hz}$) timing signal. For the $10\,\rm{m}$ prototype interferometer, well above 500 analog channels will be installed. In addition to these fast (up to $65\,\rm{kHz}$) channels, several hundred slow (a few Hertz) EPICS channels will be used; for example, environmental sensors and computers monitoring the system-wide DC power supplies provide their data via EPICS channels.
For fast communication and data sharing between real-time modules, the front-end computers are connected via a low-latency Myrinet fiber network. According to the overall status of the project, about one third of the CDS is installed and working.

\section{Laser}
\label{sec:laser}
As the sensitivity of high precision interferometry is often limited by the available laser power (i.e. shot noise) a highly stable high power laser is required. A monolithic non-planar ring oscillator (NPRO) provides a highly stable seed for a solid state amplifier. This system supplies $38 \,\rm{W}$ at a wavelength of $1064\,\rm{nm}$~\cite{laser}. While the seed is a well established, commercially available $2\,\rm{W}$ Innolight \textit{Mephisto} with Nd:YAG crystal, the amplifier is built after an LZH/AEI design for Advanced LIGO. Four Nd:YVO\textsubscript{4} rods are pumped by fiber coupled diodes at $808\,\rm{nm}$ with $150\,\rm{W}$ in total. More than $95\,\rm{\%}$ of the $1064\,\rm{nm}$ light is emitted in the TEM\textsubscript{00} mode. Mode cleaning as well as injection into vacuum are provided by a $6\,\rm{m}$ long photonic crystal fiber (type: LMA-15-PM). Its mode shape provides $\>\,99\,\%$ overlap with the fundamental Gaussian mode. The transmitted power is, however, limited to $\approx 10\,\rm{W}$ before the onset of stimulated brillouin scattering. This power level is sufficient for early experiments. Once more power is required, we plan to implement a different fiber or to use free space coupling of the light into the vacuum system at the expense of an automatic beam alignment system. A pre-mode cleaner is rigidly mounted to the freely moving optical bench inside the vacuum. It provides further spatial mode filtering and beam jitter supression and serves as fixed spatial reference for experiments.
The power fluctuations after this point are sensed to establish a feedback loop that stabilizes the output to a relative intensity noise of $5\times 10^{-9}$.
The laser is already in place and well tested while the amplitude stabilisation has been demonstrated in principle \cite{Kwee}. Coupling to the fiber is still under investigation to increase the transmitted power. Currently, about $8\,\rm{W}$ of highly stabilized light could be allocated for the sub-SQL interferometer and further experiments.
	
\section{Frequency reference}
Despite the high inherent stability of the NPRO design, a much better (seven orders of magnitude!) frequency reference is needed. The isolated tables provide a perfect environment to set up a length reference. A triangular (ring) optical cavity is formed between three suspended mirrors (see fig.~\ref{fig:layout}). The cavity round trip length will be $21.2\,\rm{m}$, the finesse of the cavity is $7300$, and it is illuminated with $130\,\rm{mW}$ of input power. To reach our requirement of $10^{-4}\,\rm{Hz}/\sqrt{\rm{Hz}}$ at $20\,\rm{Hz}$ falling with $1/f$~\cite{refc}, all three mirrors are suspended by triple cascaded pendulums isolating from lateral seismic motion while two blade spring stages yield passive vertical isolation above their corresponding eigenfrequencies of approximately $1\,\rm{Hz}$. Suspension resonances are locally sensed by BOSEM style~\cite{bosem} shadow sensors and controlled combining eddy current dampers and co-located voice coil actuators at the upper stage. The lowest stage wires are heavily loaded (30\% of breaking stress) to reduce bending losses for lowering the suspension thermal noise floor and to increase vertical bounce mode frequencies. A mirror mass of $850\,\rm{g}$ was chosen to reduce radiation pressure effects as well as substrate thermal noise. A test suspension was set up to verify that the overall weight, including surrounding cage (about $13.5\,\rm{kg}$ per suspension), is within the payload budget of the tables.

A Pound-Drever-Hall (PDH) sensing scheme is used to match the laser frequency to the length of the cavity, always maintaining resonance and thereby full sensitivity. For the chosen cavity parameters, the calculated shot noise limit of the PDH sensing is below the stability requirement. Below $1\,\rm{Hz}$ actuation is carried out by the laser temperature, up to $10\,\rm{kHz}$ a piezo actuator in the laser can be used while a phase correcting electro optic modulator (EOM) acts up to the unity gain frequency of $250\,\rm{kHz}$. Extensive simulations were carried out to design a stable controller able to achieve the required gain of up to $10^7$ within the relevant band of $20\,\rm{Hz}-100\,\rm{kHz}$. Differential wavefront sensing (DWS) and subsequent feedback control loops always overlap the ingoing beam with the cavities eigenmode by means of fast steering mirrors~\cite{refcgeom}. Static cavity misalignment is measured by spot position sensors behind the mirrors and fed to the cavity mirrors. To improve the low frequency sensitivity (i.e. below $\approx 5\,\rm{Hz}$) where passive isolation is not effective, the cavity length is stabilized to the frequency of a molecular iodine reference.

Once a second table is installed, the frequency reference cavitiy's installation can begin. Important experimental results, such as the measurement of suspension thermal noise and substrate thermal noise, might already be obtained from this setup even before the installation of the SQL interferometer, which is described in the following.
	
\section{SQL interferometer layout}
\label{sec:SQLifo}
\begin{figure*}
\resizebox{0.75\textwidth}{!}{
  \includegraphics{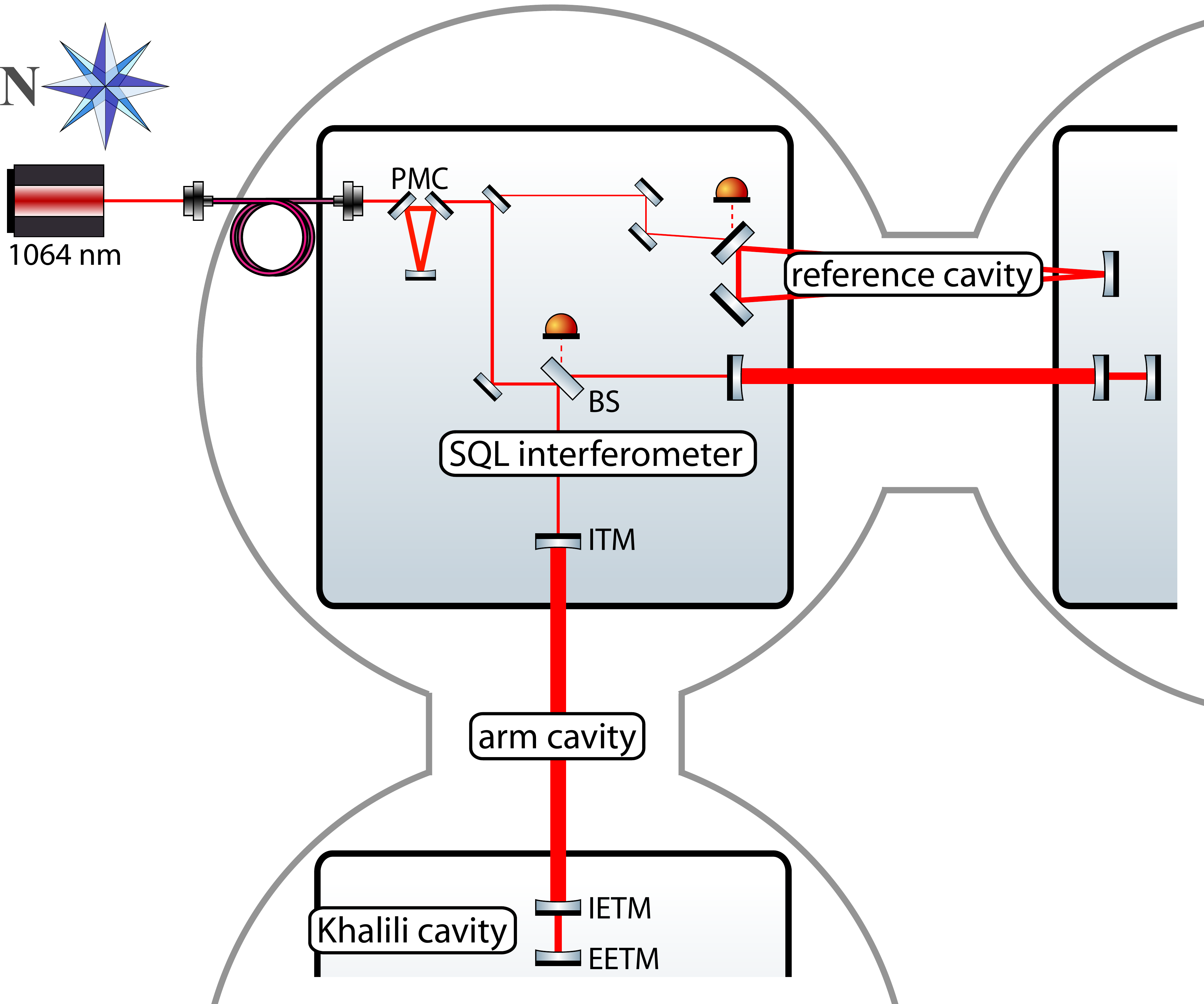}
}
\vspace*{0.5cm} 
\caption{The optical layout of the sub-SQL interferometer is shown. $8\,\rm{W}$ are sent through a photonic crystal fiber into vacuum and are additionally spatially filtered by a pre-mode cleaner (PMC). To stabilize the laser frequency, $130\,\rm{mW}$ are split of to sense the length of the triangular reference cavity with Finesse 7300. Most of the light enters the arm cavity enhanced (Finesse 670) Michelson interferometer to be sensed quantum noise limited at the antisymmetric dark tuned port.}
\label{fig:layout} 
\end{figure*}

The first major experiment designed for the AEI 10m prototype facility is a $10\,\rm{m}$ scale Michelson interferometer (see fig.~\ref{fig:layout}) optimized to be fully dominated by quantum noise in the measurement band around $200\,\rm{Hz}$. This permits interferometry at the Standard Quantum Limit equivalent to the Heisenberg limit of the $100\,\rm{g}$ interferometer mirrors, thus allowing research in macroscopic quantum mechanics.

The photon counting noise or shot noise (SN) in an interferometer read out is well investigated since it is limiting nearly all large scale interferometers such as gravitational wave detectors. The corresponding back action noise arising from the continuous position measurement with small uncertainties of interferometer mirror positions, the radiation pressure noise (RPN), could not be observed yet. Lowering the mirror mass increases the opto-mechanical coupling. The minimum possible mass was estimated to be $100\,\rm{g}$ because the fibers needed to form the suspension cannot be made thinner and their thermal noise contribution dominates the estimated mirror displacement noise. The maximum available input power is $\approx8\,\rm{W}$. All remaining parameters were designed around these values to widen the gap between classical noise and the total quantum noise to a factor of 2.5 at a SN-RPN crossover frequency of $200\,\rm{Hz}$. The baseline design foresees a Michelson interferometer enhanced by $11.4\,\rm{m}$ long arm cavities (Finesse 670) to increase the displacement signal versus SN. Mirror thermal noise will be brought down by means of large laser beam spot sizes ($r=9.7\,\rm{mm}$) on the mirrors. Coating thermal noise will be further reduced by using as few dielectric coating layers as possible. Doping the Ta\textsubscript{2}O\textsubscript{5} layers with titania reduces the loss factor by a factor of two \cite{titania}. End mirror cavities tuned to anti-resonance have the potential for further noise reduction~\cite{khalili}. In such Khalili cavities, a mirror with few double layers (e.g.~two) with low mechanical loss is used as input end test mass (IETM). Another mirror with high reflectivity (e.g.~15 layers) is used as end end test mass (EETM). This decouples loss and reflectivity. The suppression of the circulating power in the Khalili cavities compared to the arm cavities (in this case $\approx10$) reduces the read-out noise of the compound ETM by up to a factor of three. For the initial configuration no further optical recycling techniques (power-, signal recycling) are planned. The readout technique (DC or heterodyne) has not yet been decided. However, the local oscillator to enhance the signal sidebands can be split off in front of the interferometer and locked to an arbitrary readout quadrature for variational readout.

\subsection{Optics suspensions}
	\label{suspension}
	Although a well isolated platform is available with the seismically isolated tables, the investigation at the border of the quantum world requires much better ($\approx$ eight orders of magnitude) isolation. Therefore all relevant mirrors will be suspended as multiple stage pendulums with special emphasis on the minimization of thermal noise. As with the reference cavity mirrors, all interferometer mirrors will be suspended in triple cascaded pendulum stages, isolating horizontal seismic noise. Both upper stages will be mounted from blade springs for vertical isolation. Slow actuation and local damping will take place at the uppermost stage using BOSEMs. To provide fast actuation newly developed electrostatic drives are planned at the lowest stage. Due to the small mirror size (diameter $50\,\rm{mm}$), former suspension construction experience must be mostly redefined. A special challenge is to lower the resonance frequencies of all degrees of freedom not to spoil the measurement band of $20\,\rm{Hz}$ to $1\,\rm{kHz}$, especially keeping longitudinal resonances low to benefit from the $1/f^2$ transmissibility of the suspension. Thermal motion can be limited by minimizing loss factors. Therefore the lowest stage will be suspended monolithically using thin fused silica fibers of $20\,\rm{\mu m}$ diameter, the thinnest reproducible high-quality fibers currently available. These fibers will be loaded to $30\,\%$ of their breaking strength to reduce the energy that is stored in the fiber bending and hence increase dilution.

\subsection{Reaching the design sensitivity}
Large beam spots on the arm cavity mirrors come at the expense of operating both the arm cavities as well as the "Khalili cavities" close to instability ($g_{\rm{arm}}\approx0.99$, $g_{\rm{khalili}}\approx0.99998$). To prevent inoperability of the interferometer, a stepwise approach is planned \cite{Christian}. Initially, the Khalili cavities are omitted to reduce the length degrees of freedom to be controlled from five to three. Furthermore, the arm cavities are set up shorter reducing the cavity g-factor at the expense of a smaller beam size. Therefore, this initial configuration will be limited by coating thermal noise between 100\,Hz and 1\,kHz. Once realized, the arm length can be increased stepwise by moving the end mirrors towards their desired position. In a second step, Khalili cavities will be employed to reach the interferometer's design sensitivity.
\begin{figure*}
\resizebox{0.75\textwidth}{!}{
  \includegraphics{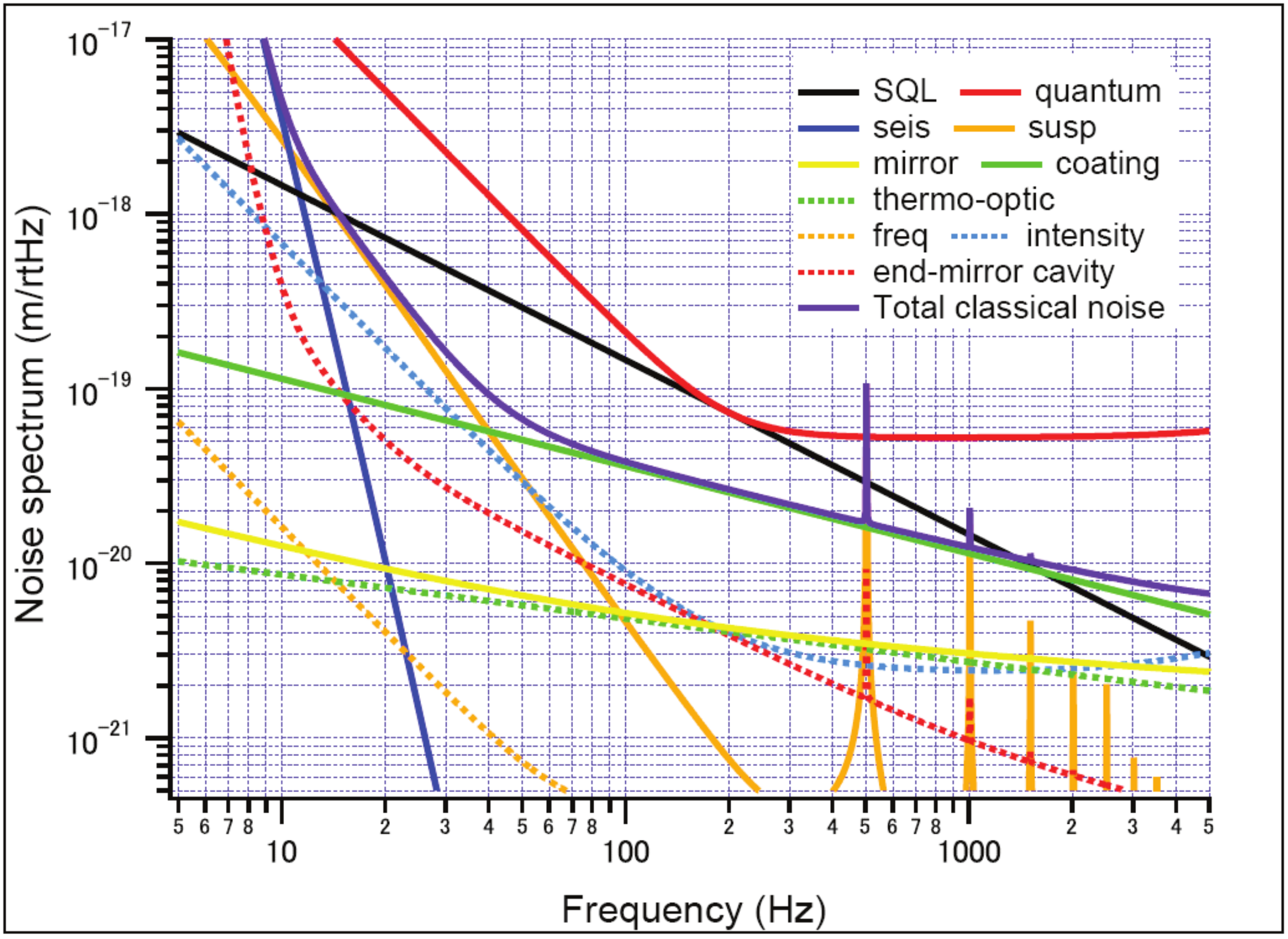}
}
\vspace*{0.5cm}       
\caption{The design displacement sensitivity of the SQL interferometer is optimized to be purely quantum noise limited over a wide frequency range, i.e. photon shot noise at high frequencies ($\ge 200\,\rm{Hz}$) and quantum radiation pressure noise at lower frequencies. The margin of 2.5 above the total classical noise dominated by coating thermal noise at the crossover frequency and suspension thermal and seismic noise at the low frequency end enables the investigation of Heisenberg limited behavior of macroscopic ($100\,\rm{g}$) test masses. Furthermore techniques of quantum noise manipulation such as squeezing injection or quantum dense readout can be tested.}
\label{fig:sens}    
\end{figure*}

\section{Conclusion}
With the techniques described, the total technical noise can be kept a factor of 2.5 below the quantum noise (see fig.~\ref{fig:sens}). Once having observed RPN and the SQL, overcoming them by non-classical interferometry using injection of (frequency dependent) squeezing, variational readout or signal recycling, will open completely new possibilities. Further down the road with a similar setup even entanglement of macroscopic objects (test mass mirrors) and the investigation of gravity self decoherence might come into reach.

\begin{acknowledgement}
This work was supported by the QUEST cluster of excellence of the Leibniz Universit\"at Hannover.
\end{acknowledgement}

\end{document}